\documentclass{iau-temp}
\pdfoutput=1 
\usepackage{amsmath}
\usepackage{graphicx}
\usepackage{multirow}

\begin{document}

\lefttitle{A.Lawrence}  
\righttitle{Astronomy, Doughnuts, and Carrying Capacity} 

\jnlPage{1}{7}
\jnlDoiYr{2021}
\doival{10.1017/xxxxx}

\aopheadtitle{Proceedings IAU Symposium}
\editors{C. Walker,  D.Turnshek, P.Grimley, D.Galadi-Enriquez \&  M.Aub\'e, eds.}

\title{Astronomy, Doughnuts, and Carrying Capacity}

\author{A.Lawrence}
\affiliation{Institute for Astronomy, University of Edinburgh}

\begin{abstract}
I examine the applicability of ecological concepts in discussing issues related to space environmentalism. Terms such as ``ecosystem'', ``carrying capacity'', and ``tipping point'' are either ambiguous or  well defined but not applicable to orbital space and its contents; using such terms uncritically may cause more confusion than enlightenment. On the other hand, it may well be fruitful to adopt the approach of the Planetary Boundaries Framework, defining trackable metrics that capture the damage to the space environment.  I argue that the key metric is simply the \textit{number} of Anthropogenic Space Objects (ASOs), rather than for example their reflectivity, which is currently doubling every 1.7 years; we are heading towards degree scale separation. Overcrowding of the sky is a problem astronomers and satellite operators have in common. 
\end{abstract}


\maketitle

\section{Introduction}

An article in Nature Astronomy in 2022 put forward the scientific and technical case for ``Space Environmentalism'', a concept that had been developing informally for some time \citep{Lawrence_case_22}. I still feel that ``Space Environmentalism" is a better term than ``Space Sustainability" as, with the latter term, it is not clear what we are trying to sustain, and the term ``sustainability" is slowly gathering an aura of greenwashing. An environmental approach captures several key elements - that actions may have damaging consequences elsewhere in the environment; that the true cost of an activity is displaced onto others occupying the same environment; and that damage is incremental and complex. All those aspects mean that we should take a broad environmental approach to our problems rather than a narrow self-enclosed one, and encourage a sense of stewardship of an environment shared by astronomers, by space operators, and by the general public, especially amateur astronomers and indigenous communities. However, \cite{Lawrence_case_22} also suggested in passing that space could perhaps be seen as an \textit{ecosystem}, and many discussions since have used concepts carried over from ecology, such as the idea of carrying capacity.  In this paper I want to critically examine whether ecological concepts can be sensibly applied to orbital space, and if not, what we should use instead.

\section{Applicability of of ecological concepts to orbital space.}

Much of what I refer to below comes from standard ecology. A popular ecology textbook is \cite{Chapin_book}, and a very practical one for astronomers new to these topics (like me!)  is \cite{Lehman_book}. 

\subsection{Is orbital space an ecosystem?}

Ecosystems are normally defined in terms of organisms which interact with each other, and which also interact with their physical environment (nutrients, energy flows, etc). But even if we generalise to more general abiotic complex systems, to usefully think of such a system as an ecosystem we need feedback loops between the components, in which case orbital space and its contents does not look like an ecosystem. (See \cite{Thomson_23} for more detail.)
On the other hand, if we combine the contents of orbital space, the human actors on Earth that launch into space, and the economic and regulatory system they find themselves in, this broader system does look like a human-economic ecosystem. However, it's not clear how this insight helps.

\subsection{How do we define the ``carrying capacity" of orbital space?}

Some papers have unpacked the historical development of the term ``carrying capacity"  and noted how inconsistently and ambiguously it is used \citep{Sayre_08, Chapman_18}. It is possible to discern three distinct concepts for which the term ``carrying capacity" has been used. (i) The carrying capacity of a train is just the number of people that can fit on before it is full. This isn't a concept that applies to orbital space. (ii) The carrying capacity of a cargo ship is how much you can safely load it with, before it sinks. This seems to be what people sometimes have in mind when talking loosely about carrying capacity in orbital space - we mustn't exceed the carrying capacity or there will be disaster! This "safety level" meaning doesn't obviously apply to orbital space. (iii) In population ecology, the term ``carrying capacity" has a very well defined technical meaning. It is an \textit{equilibrium level} which emerges naturally from the behaviour of the Van der Hulst equation used to model the time evolution of a population with density dependent effects. If the population is below this level, it will rise, and if it is above it will fall back. Implicit in this equation are a death rate and a birth rate which depend on some kind of renewable food resource. One can also model competing populations using the Lottke Volterra equations; each population will find its own equilibrium level, including the possibility of one population wiping out the other. But none of this obviously applies simply to the behaviour of orbiting satellites. 

\cite{Sturza_22}  carry out some very interesting work in which they define ``carrying capacity'' to be the point at which the rate that satellites are consumed by collisions is equal to a specified fraction of the constant rate at which satellites are launched to maintain a constellation. This is related to the ``tipping point'' concept which we discuss below, and is important and impressive work, but I think that using the term ``carrying capacity'' in this context is potentially misleading and confusing.

\subsection{Do we expect a Tipping Point?}

Standard population ecology models assume the availability of environmental resources (such as food) which are renewable. A population or quantity embedded in the environment may saturate, or it may grow indefinitely, depending on the density-dependence of the birth and death rates. If however the system includes a \textit{non-renewable resource}, the population can rise to a maximum and then decline. This was the central insight of the Club of Rome ``Limits to Growth" work \citep{Meadows_72,Meadows_05}.  In the context of population ecology models, a standard Van der Hulst style carrying capacity could be defined at any one instant, but it would change with epoch, and so could be seen as a ``dynamic carrying capacity". This idea is implicit in the qualitative illustrations of Fig. 3 of \cite{Miraux_22}. There is however no obvious non-renewable resource that is being used up in orbital space.

Overall we have to be very cautious about the direct transference of concepts from mathematical ecology. Naively using concepts like ``carrying capacity" or ``tipping point" is likely to cause more confusion than enlightenment. However, this is not to say that we cannot model the time evolution of the contents of orbital space with simple rate equations embodying things such as launch rate, satellite lifetime, collision probability, and so on. This is exactly what what was done with the JASON debris model \citep{Long_21, Lawrence_case_22} and in the work of \cite{Sturza_22}. 

Fig. 6 in \cite{Lawrence_case_22} shows an example of a model where too large a launch rate into a particular orbital shell is counter productive in the long term - the gradual accumulation of debris in that shell slowly increases the probability of disabling collisions, and the number of net active satellites starts to go down. However, this process takes several decades, it is not a sudden disaster, and it applies to a specific orbital shell. As we will see below, there are more urgent  things to worry about.

\section{The planetary boundaries framework and doughnut economics}

An interesting suggestion made by both \cite{Miraux_22} and \cite{Wilson_23}
is that space environmentalism issues could be encapsulated in the Planetary Boundaries Framework 
\citep{Rockstrom_09}. In this framework, a variety of metrics are defined that can be used to monitor the health of the planet - for example, ocean acidification, CO$_2$ concentration, stratospheric ozone depletion, and so on. In the most recent report, \cite{Richardson_23} argue that the planet is beyond safe boundaries in 6 out of their 9 metrics. A tenth boundary is left for ``novel entities". This could perhaps be used for sky brightness \citep{Miraux_22} or ``orbital space depletion" \citep{Wilson_23} but it could also be some metric more closely related to optical or radio interference (see section \ref{sky_density}). It would be very fruitful to come to an agreed community definition of what could be used.

As also noted by \cite{Wilson_23} the well known ``Doughnut Economics" model of 
\cite{Raworth_17}  adds inner boundaries to the planetary boundaries concept. Humanity needs enough activity to produce sufficient food, energy, health, social equity and so on, but not so much that we exceed the safe outer planetary boundaries - we would like to be inside the doughnut, the  ``safe and just space for humanity". This is a good way to look at the tension between astronomy and space industry that we currently face - we want activity that builds towards goals such as connectivity for all, but without causing too much damage to the sky, or creating so much debris collision risk that space industry becomes unsustainable. But how do we define and agree how much is enough?

One aspect of looking at space environmentalism as an economic issue, and a key issue in Raworth's work, is challenging the assumption that growth is always good. We should be aiming for ``thriving" rather than growth \textit{per se}. As physicists, we are aware that a constant rate of growth leads to an exponential. Qualititatively, the population of anthropogenic space objects (ASOs) seems to be accelerating ... does it in fact show exponential growth?

\section{Exponential growth of space activity}

Fig. 1a shows the log of the number of currently active satellites versus linear time, using data kindly provided by J.McDowell, extracted from the General Catalog of Artificial Space Objects \citep{McDowell_23}. On such a plot, exponential growth will be of course a straight line. What we see is three distinct historical phases, each showing exponential growth, at different rates. In the original Space Age, the number of active satellites was doubling every 1.2 years. We then have several decades of Normal Space Activity, dominated by national space agencies and military satellites, but with of course some commercial communications satellites. In this era, space activity was growing much more slowly, doubling every 16.7 years.  Finally we reach the New Space era, with activity recently doubling every 1.7 years. This is currently dominated by the Starlink constellation of course, but we can see that the New Space era in fact began before Starlink. With Starlink excluded, we see a doubling time of  5.0 years, starting around 2015. Then the Starlink project begins a second kick upwards from 2019 onwards. Without having to guess the fate of various proposals before the FCC etc, one could predict the future population empirically by extrapolation.

\begin{figure}[t]
\centering
  \centerline{\vbox to 6pc{\hbox to 10pc{}}}
  \includegraphics [width=0.95\textwidth]  {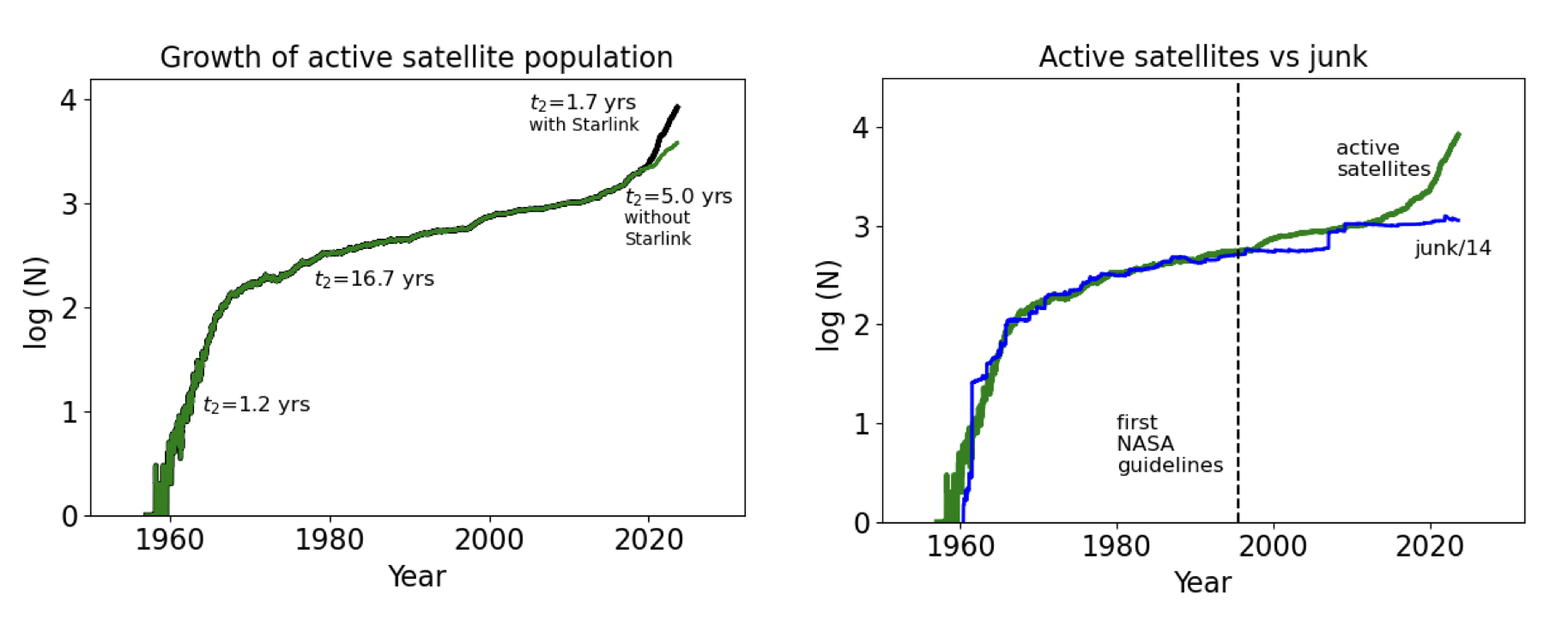}
  \caption{(a) Left: Growth \textit{vs} time of the active satellite population, in log-linear form, using data from \cite{McDowell_23}. Approximate doubling times of the three apparent historical segments are marked. (b) Right: active population compared to the ``junk'' population, divided by 14. See text for details.}
  \label{Fig1}
\end{figure}

How has the space junk population grown? Here we define ``junk" as the total of tracked debris and various ``leftovers" such as rocket bodies, defunct satellites and so on. Once again, the data comes from the General Catalog of Artificial Space Objects \citep{McDowell_23}. In the original Space Age, junk grew even faster than the active population, and then slowed its growth to a rate similar to the active population. Fig. 1b shows the number of junk objects divided by 14, to (roughly) match the active population. After the very rapid initial rise, the junk curve closely follows the active curve until about 1995. Roughly speaking each active object is associated on average with 14 pieces of  junk. (Note that each curve represents the number of objects extant at that time, i.e. does not include objects that had already re-entered.)

After about 1995, the junk curve flattens off and grows much more slowly than the active population. In particular, the recent dramatic growth of active objects is not matched by a similar growth in junk. The year 1995 is when NASA first issued its orbital debris mitigation deadlines which were then adapted and improved over the next few years by the Inter Agency Debris Co-ordination committee (IADC). (See the history summarised at \href{https://orbitaldebris.jsc.nasa.gov/mitigation/}{https://orbitaldebris.jsc.nasa.gov/mitigation/}.) It seems that \textit{debris mitigation works}.

A large fraction of the junk growth that has been made is due to either the Irridium-Kosmos collision of 2009, or the Chinese ASAT of 2007, which even on this log plot are clearly visible as vertical jumps. On the other hand, the debris caused by the 2021 Russian ASAT is visible only as a temporary blip; at the low altitude concerned, most of the debris made has now re-entered. And of course a major reason why New Space activity has not produced a major burst in net debris is because nearly all this growth has been in LEO, and indeed the Starlink project deliberately aims at a \textit{turnover} of population, constantly demising and replacing objects. 

Our broad-brush approach here of course masks the fact that the collision risk problem is about specific crowded orbital shells. Also, although some mitigation is achieved partly by ``parking" objects in higher orbits after the 25 year lifetime guideline, it is mostly achieved by atmospheric re-entry. The consequence is that junk is now more than ever raining down upon us. We know that burn-up is often incomplete, with large pieces landing on the surface of the Earth, and many more smaller pieces causing an ever increasing risk to air traffic (see \cite{Lawrence_case_22}. For this reason, and many others, it seems urgent that \textit{Third Party Liability} insurance should become mandatory, just as it is for motor vehicles in most countries. 

\section{Sky density of active objects}
\label{sky_density}

Most mitigation discussions between astronomers and space operators centre on satellite reflectivity, where some progress has been made (at the same time as satellites are getting bigger!). Astronomers also have opinions about orbital height, but this is a bit more subtle, with tradeoffs between hours in sunlight, brightness, and angular speed. Of course from the radio point of view, satellites are visible whenever they are above the horizon. We all know however that the real problem is the \textit{number} of satellites. What is the point of making satellites half as bright if there are fifty times as many of them? Furthermore, large modern telescopes can potentially see all the satellites from here to GEO - as long as they are above the horizon - so to a first approximation orbit height doesn't matter. In reality, the fraction of the satellites in a particular orbital shell that are visible above the horizon, and where they are on the sky at a given time of night, depends on the height of the shell, the mixture of orbital inclinations, and the latitude of observation. However we can obtain a simple metric that captures the growth of sky density by ignoring those subtleties. Fig. 2a shows the growth of this metric, assuming a random distribution of orbit inclinations and placing all satellites at $h=500$ km. (Using a low height will tend to underestimate the number in the sky).  In the 1970s there were typically a handful of satellites in the sky at any one time; today it is over a hundred. (Note that only some of these are visible to the naked eye). Fig. 2b shows the corresponding decline in the typical distance between two satellites. We are heading towards degree scale separation.

\begin{figure}[t]
 \centering
  \centerline{\vbox to 6pc{\hbox to 10pc{}}}
  \includegraphics[width=0.95\textwidth] {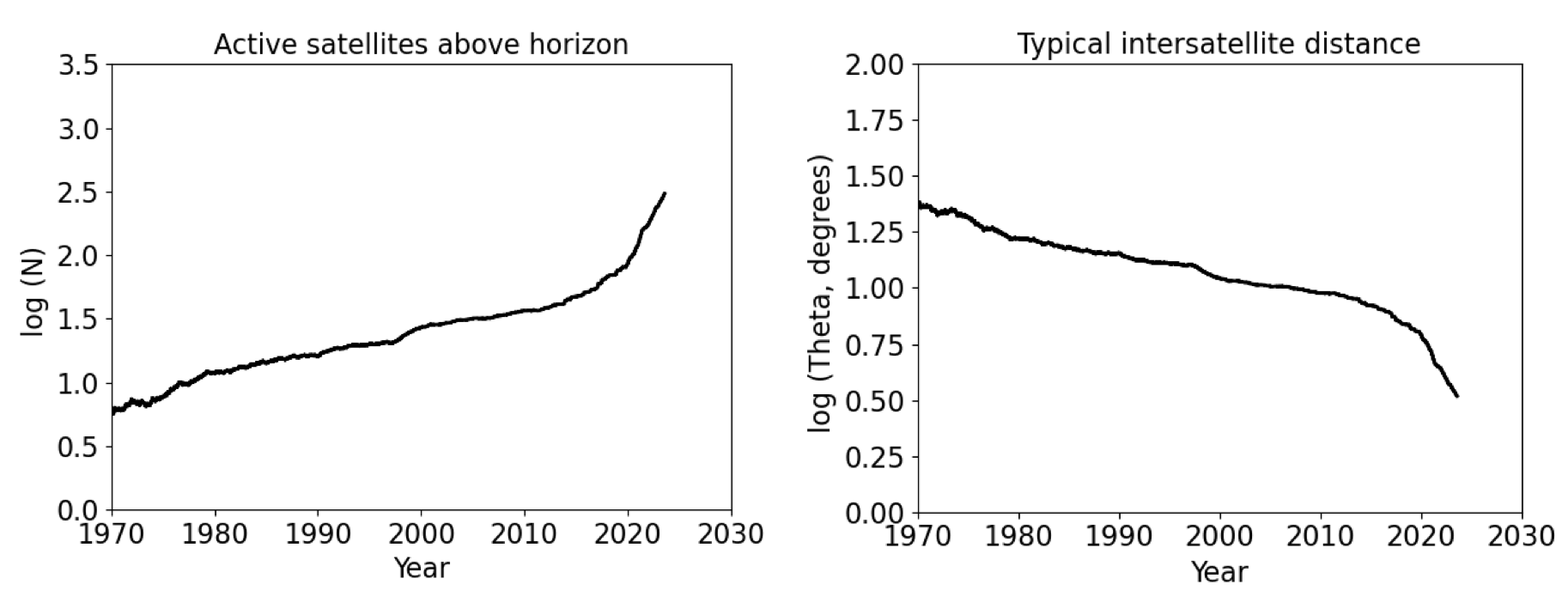}
  \caption{(a) Left: Historical growth of the approximate average number of satellites above the horizon, using a very simplified height and orbit model and data from \cite{McDowell_23}. See text for details. (b) Right: the corresponding typical inter-satellite distance on the sky.}
  \label{Fig2}
\end{figure}

Satellite operators are also affected by the inexorable increase in satellite communications overcrowding, as spelled out by in a recent Viasat White Paper \citep{Viasat_22}
There are two problems, related to the optical and radio astronomy interference problems. The first problem is the increase in ``line of sight collisions", where for example a ground station trying to communicate with a satellite in GEO finds a LEO satellite passing across. For many years, it has been agreed that GEO satellites should be at least one degree apart on the sky, leading to a finite number of highly competitive GEO slots. Of course, LEO satellites are constantly moving, so what we have now is an equivalent statistical problem. This is closely analogous to the problem of satellites every so often streaking across a wide field astronomy optical image. The second issue is the aggregate power problem, from the sidelobes of the many satellites over the sky. This is of course closely analogous to the problem of radio astronomy interference. This problem includes unintentional out-of-band emission, just as radio observatories are now finding. (See elsewhere in these proceedings.)  

It seems to me that this problem - sky density communications overcrowding - is going to hit crisis point long before the Kessler syndrome. The effectively useable sky may be captured by a small number of operators, to the detriment of other operators, and astronomers. 

\section{Conclusions}

My main message is: \textbf{The problem is the numbers not the brightness}. We should not lose sight of the real problem. But in order to fight that battle, we need a defensible position on this question: \textbf{How many is enough?}.  Meanwhile, here are some more detailed conclusions: \\

\begin{itemize}
\item We should be very cautious about using ecological concepts uncritically
\item It could be useful to see space environmentalism in the planetary boundaries framework
\item Satellite population growth shows three distinct historical exponential phases
\item The current doubling time is roughly two years
\item Space junk growth slowed from 1995 onwards
\item Debris mitigation guidelines seem to work
\item Material must be descending through the atmosphere at an increasing rate
\item We need a new collision liability framework
\item The satellite population is approaching degree-scale separation
\item Operators have the same interference problems that we do!
\end{itemize}


\end{document}